\documentclass[pra,letterpaper,onecolumn,showpacs,superscriptaddress]{revtex4}
\usepackage{graphicx,psfrag,amsmath,amssymb,amsfonts,bbm,latexsym,color,dcolumn}

\begin{document}

\title{Uncertainty-principle noise in vacuum-tunneling transducers}

\author{Carlo Presilla}

\affiliation{Dipartimento di Fisica, Universit\`a di Perugia, Perugia
  06100, Italy}

\author{Roberto Onofrio}

\affiliation{Dipartimento di Fisica, Universit\`a di Roma ``La
  Sapienza'', Piazzale A. Moro 2, 00185 Rome, Italy}

\affiliation{Department of Electrical Engineering, University of
  Rochester, Rochester, NY 14627}

\author{Mark F. Bocko}

\affiliation{Department of Electrical Engineering, University of
  Rochester, Rochester, NY 14627}

\date{15 February 1992, published in {\sl Phys. Rev. B} {\bf 45} (1992) 3735}

\begin{abstract}
The fundamental sources of noise in a vacuum-tunneling probe used as
an electromechanical transducer to monitor the location of a test mass
are examined using a first-quantization formalism. We show that a
tunneling transducer enforces the Heisenberg uncertainty principle for
the position and momentum of a test mass monitored by the transducer
through the presence of two sources of noise: the shot noise of the
tunneling current and the momentum fluctuations transferred by the
tunneling electrons to the test mass. We analyze a number of cases
including symmetric and asymmetric rectangular potential barriers and
a barrier in which there is a constant electric field. Practical
configurations for reaching the quantum limit in measurements of the
position of macroscopic bodies with such a class of transducers are studied.
\end{abstract}

\maketitle

\section{Introduction}

The investigation of the ultimate quantum limits for the detection of
weak forces has been stimulated by the development of sensitive
antennae to search for gravitational wave radiation \cite{PRB1,PRB2}. 
If several practical barriers can be overcome and electromechanical
transducers of sufficient sensitivity are developed, then it should be
possible to monitor massive Weber-bar gravitational wave antennae in
the regime in which their behavior is dominated by quantum effects,
{\it i.e.}, by the measurement process itself. Thus a class of
experiments in which repeated measurements are performed on a single
isolated macroscopic quantum-mechanical oscillator may become possible
\cite{PRB3,PRB4}.

So far superconducting-quantum-interference-device (SQUID)-based
electromechanical transducers have offered the best opportunity to
study the quantum regime. However, recently it was pointed out that
the tunneling probe used in the scanning tunneling microscope is a
quantum limited electromechanical amplifier and therefore may present
an opportunity to study the quantum regime with electromechanical
transducers \cite{PRB5}. 

Since the tunneling transducer is intrinsically a quantum device,
without a classical analog, a quantum analysis is required to
understand the origin of its noise. It was shown that there are two
independent sources of noise in the tunneling transducer
\cite{PRB6,PRB7}. The first is the well-known shot noise of the
tunneling current which enters as an apparent fluctuation of the test
mass. The other source of noise is a fluctuating ``back-action'' force
which the tunneling transducer exerts on the test mass. The two
sources of noise work in concert to add to the amplified mechanical
signal an amount of noise power equivalent to one-half quantum of
energy per second at the operating frequency. Recently Yurke and
Kochanski presented a full quantum-mechanical analysis of the noise of
a tunneling transducer \cite{PRB8}. They used a second-quantized
description of electron tunneling through a barrier to find an
expression for the uncertainty in the width of the tunneling barrier,
which is equivalent to the position of the test mass, based upon the
tunneling current fluctuations. They also computed the fluctuation of
the momentum current transported across the barrier. Their
calculations explicitly show that the tunneling transducer enforces
the Heisenberg uncertainty relation between the position and momentum
of the test mass. 

The purpose of this paper is twofold. The first purpose is to present
a simplified, first-quantization treatment of the noise in the
tunneling transducer. Although we obtain the same expressions for the
uncertainties as Yurke and Kochanski in Ref. 8, we think that the use
of first quantization to deal with this problem is more physically
intuitive and less mathematically complex. The second purpose of this
paper is to discuss some of the practical considerations regarding the
tunneling transducer and the prospects for achieving quantum noise
limited force detection.

The paper is organized as follows. In Sec. II, after a brief
description of the working principles of the tunneling transducer, we
express the position and the momentum uncertainties in terms of the
time-independent solutions of the Schr\"odinger equation. The position
uncertainty is derived from the transmission coefficient and the
momentum uncertainty is obtained by a generalization of the current
flux. In Sec. III we apply these considerations to calculate the
position and momentum uncertainty product for symmetric and asymmetric
rectangular barriers. In Sec. IV we treat the case of a barrier in
which there is a constant electric field. In Sec. V we discuss the
practical obstacles to achieving quantum noise dominance and we give a
specific example of a configuration in which quantum effects may be
observed. In Sec. VI we discuss some conceptual problems in making the
correspondence between the quantum mechanical uncertainties which are
calculated here and the more experimentally relevant classical
description of noise which employs spectral densities of random variables.
 
\section{Position and momentum uncertainties for a tunneling
  transducer}

The tunneling transducer is simply a variable resistance
transducer. The motion of a test mass modulates the gap of a vacuum
tunnel junction thus affecting the tunneling probability. If the
junction is voltage biased then the current measured by an amplifier
which follows the tunnel probe provides a sensitive measure of the
tunneling gap and therefore of the displacement of the test mass. 

In Figure 1 we show a schematic representation of the tunneling
transducer. A tunneling tip is places a distance $l$ from the test
mass which is going to be monitored. The displacement of the test mass
from its initial position is given by $x$. The effective resistance of
the tunneling transducer is given, in the limit $k_0 l \gg 1$, by the
familiar formula

\begin{equation}
R=R_0 e^{-2k_0 x},
\end{equation}  
where $k_0$ is the inverse of the de Broglie wavelength of the
electrons with energy $E$ inside the barrier of height $V_0$ and is
given by

\begin{equation}
k_0=\left[ \frac{2m(V_0-E)}{\hbar^2} \right]^{1/2}.
\end{equation}
The tunneling resistance $R_0$ is usually around $10^6-10^8 \Omega$
and a typical value of $k_0$ is $10^{10}$ m${}^{-1}$; the distance
scale over which the tunneling resistance change significantly is of
atomic dimensions. For typical values of the tunnel probe voltage bias
the tunneling current is in the range of nanoamps to microamps. Using
conventional electronic techniques it is possible to measure extremely
small fractional changes in currents of this magnitude so it is
possible with the tunnel junction transducer to measure displacements
which are a very small fraction of $k_0^{-1}$.

It is important that the capacitance between the tunneling probe and
the test mass be small, on the order of $10^{-17}$ F or less \cite{PRB7}.
This ensures that the quantum effects associated with the tunneling
transducer will dominate the back-action force fluctuations that have
their origin in the amplifier used to sense the tunneling current and
which are capacitively coupled to the test mass. This assumption
allows us to concentrate on the fluctuations which arise from the
tunneling process.

We express the uncertainties in the position and the momentum of the
test mass which is sensed by the tunneling transducer in terms of
solutions of the time-independent Schr\"odinger equation which
describes the motion of a particle in the presence of a
one-dimensional barrier. Let us assume that there are $N$ electrons
attempting to tunnel out of the probe. We treat each tunneling process
as independent from the others, this approximation being satisfactory
for the realistic tunneling currents that can be obtained. For each
electron there is a probability $T$ that it will tunnel and a
probability $R=1-T$ that it will not tunnel, where $T$ and $R$ are,
respectively, the transmission and the reflection coefficients
associated with the barrier. The probability that $n < N$ electrons
will escape from the probe is given by the binomial law, therefore the
average number of electrons which escape will be $\langle n
\rangle=NT$ and the variance of the average is 

\begin{equation}
(\Delta n)^2=\langle (n-\langle n \rangle)^2\rangle=NTR
\end{equation}

The variance of the number of electrons which tunnel may be written as
a function of the transmission coefficient and the gap between the tip
and the test mass:

\begin{equation}
\Delta n=N |\frac{\partial T}{\partial l}| \Delta l
\end{equation}
and the uncertainty in the position of the barrier therefore is
inferred as

\begin{equation}
\Delta l=\frac{1}{\sqrt{N}} \frac{\sqrt{TR}}{|\frac{\partial
    T}{\partial l}|}.
\label{2.5}
\end{equation}

\begin{figure}[t]
\includegraphics[clip,width=0.40\columnwidth]{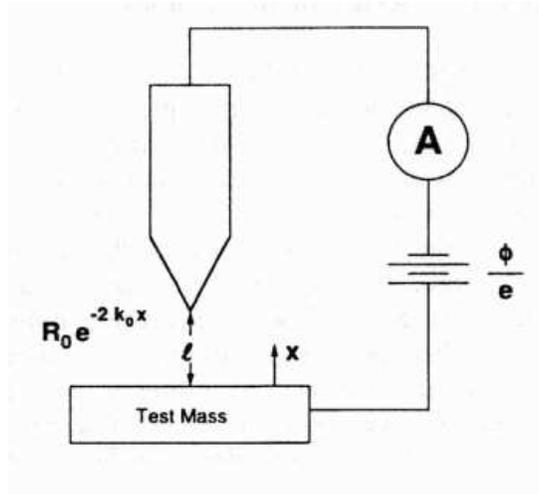}
\caption{Scheme for the detection of displacement through a
  vacuum-tunneling transducer. The ``at-rest'' separation between the
  test mass and the tunneling probe is $l$ and the displacement of
  the test mass is $x$. The probe is voltage biased and the current is
  sensed by a current amplifier $A$.}
\label{fig1}
\end{figure}

When $|\partial T/\partial l|=0$ a second-order expansion must be
employed; however, in all the situations which we explore in what
follows, a first-order expansion of (\ref{2.5}) is adequate.

In order to calculate the uncertainty in the momentum transferred to
the barrier we first consider the continuity equation for the
probability flux

\begin{equation}
\frac{\partial \rho}{\partial t}+\frac{\partial J}{\partial x}=0
\label{2.6}
\end{equation}
where the probability density is 

\begin{equation}
\rho=\psi^* \psi
\end{equation}
and the probability current is

\begin{equation}
J=-i \frac{\hbar}{2m}
\left[\psi^* \frac{\partial \psi}{\partial x}-
      \frac{\partial \psi^*}{\partial x} \psi \right].
\end{equation}
Analogously for the momentum flux we have the following conservation
equation:

\begin{equation}
\frac{\partial \rho}{\partial t}+\frac{\partial J_p}{\partial
  x}=-\frac{\partial V}{\partial x} \psi^* \psi,
\label{2.9}
\end{equation}
where the momentum density is

\begin{equation}
\rho_p=-i \frac{\hbar}{2} \left[ \psi^* \frac{\partial \psi}{\partial
    x}-\frac{\partial \psi^*}{\partial x}\psi \right].
\end{equation}
and the momentum current is given by

\begin{equation}
J_p=\frac{\hbar^2}{4m} \left[ 2 
\frac{\partial \psi^*}{\partial x} \frac{\partial \psi}{\partial x}-
\psi^* \frac{\partial^2 \psi}{\partial x^2}-
\frac{\partial^2 \psi^*}{\partial x^2} \psi \right].
\label{2.11}
\end{equation}

This progression can be carried out to higher moments of the momentum
and for our calculation of the variance of the momentum we need to
consider the flux of ``momentum squared'' for which the following
continuity equation applies:

\begin{equation}
\frac{\partial \rho_{p^2}}{\partial t}+
\frac{\partial J_{p^2}}{\partial x}=i \hbar \frac{\partial V}{\partial
  x}  \left[ \psi^* \frac{\partial \psi}{\partial
    x}-\frac{\partial \psi^*}{\partial x}\psi \right]
\label{2.12}
\end{equation}
in which
\begin{equation}
\rho_{p^2}=-\frac{\hbar^2}{2} \left[ \psi^* \frac{\partial^2 \psi}{\partial
    x^2}-\frac{\partial^2 \psi^*}{\partial x^2}\psi \right]
\end{equation}
and

\begin{equation}
J_{p^2}=i \frac{\hbar^3}{4m}
\left[\psi^* \frac{\partial^3 \psi}{\partial x^3}-\frac{\partial
    \psi^*}{\partial x} \frac{\partial^2 \psi}{\partial x^2}+
\frac{\partial^2 \psi^*}{\partial x^2} \frac{\partial \psi}{\partial
  x} -\frac{\partial^3 \psi^*}{\partial x^3}\psi \right].  
\label{2.14}
\end{equation}

The above equations were derived by forming combinations of successive
derivatives of the time-dependent one-dimensional Schr\"odinger
equation for a particle in a potential $V(x)$ in much the same way as
the familiar continuity equation (\ref{2.6}) is derived. Note that (\ref{2.9})
expresses Newton's law in quantum mechanical terms, the right-hand
side of (\ref{2.9}) being the force density which acts on the particle. 
Also, a feature of (\ref{2.14}) deserves comment. As we will see in the
following examples $J_{p^2}$ is negative inside the barrier which is a
consequence of the following. The barrier is a classically forbidden
region so the kinetic energy flux $J_{p^2}/2m$ associated with a
particle inside the barrier is negative which makes the ``momentum
squared'' flux $J_{p^2}$ negative also. 

Following Yurke and Kochanski's approach let us imagine that the
potential $V(x)$ represents a barrier located between $a<x<b$, and 
$V(x)$ is zero outside of this region. We decompose the force of the
potential barrier on a tunneling particle into two parts, $\partial
V/\partial x=\partial V_1/\partial x+\partial V_2/\partial x$, where
$V_1$ is associated with the tunneling probe at the location $a$ and
the potential $V_2$ is attributed to the test mass surface at $b$. To
calculate the momentum uncertainty imparted to the test mass we must
find the momentum and 'momentum-squared'' fluxes passing through a
surface at $b$. The momentum current transferred to the part of the
barrier at $b$ is obtained by using the stationary version of (\ref{2.9})
where $V(x)$ is replaced with $V_2(x)$, {\it i.e.}
\begin{equation}
J_p^t=J_p(b^+)+\int_{a^-}^{b^+} \frac{\partial V_2}{\partial x} \psi^*
\psi dx.
\label{2.15}
\end{equation} 
The ``momentum squared'' flux transferred to the barrier at $b$ is,
using Eq. (\ref{2.12}),
\begin{equation}
J_{p^2}^t=J_{p^2}(b^+)-i \hbar \int_{a^-}^{b^+} \frac{\partial
  V_2}{\partial x} \left[\psi^* \frac{\partial \psi}{\partial x}-
\frac{\partial \psi^*}{\partial x} \psi\right] dx
\label{2.16}
\end{equation}
Dividing $J_p^t$ and $J_{p^2}^t$ by the incident flux $J_{\mathrm in}$
we obtain the momentum and ``momentum squared'' transferred to the
potential barrier at $b$ by a single tunneling particle. Finally, the
momentum and ``momentum squared'' transferred to the test mass is
related in the following fashion to the momentum and ``momentum
squared'' transferred to the barrier from the electron. In our model,
the test mass is schematized by a potential step at a fixed location
in space. The test mass can be thought of as an infinitely rigid
oscillator with a surface fixed at the location $b$; this is the
effect of the feedback system which is actually used to prevent the
test mass position from drifting under the effect of the continuous
stream of electrons impinging on it. Therefore the test mass has a
quantum-mechanical wave function which rapidly decays away from $b$. In
a plane-wave representation, this corresponds to a superposition of
plane waves with imaginary momenta. Thus a localized particle, in this
case the test mass at $b$, with real momentum can be viewed as a free
quasiparticle with imaginary momentum. Thus the mean momentum imparted
to the test mass is $\langle p \rangle=i J_p^t/J_{in}$ which is $i$
times the momentum transferred to the barrier from each electron
tunneling event. As a consequence the mean ``momentum squared''
imparted to the test mass is $\langle p^2 \rangle=-J_{p^2}^t/J_{in}$. 
The momentum fluctuation of the test mass due to $N$ electrons is then
\begin{equation}
(\Delta p)^2=N(\langle p^2 \rangle-\langle p \rangle^2)=N \left[-
\frac{J_{p^2}^t}{J_{in}}+\left(\frac{J_p^t}{J_{in}}\right)^2\right].
\label{2.17}
\end{equation}

In the following sections we will calculate the uncertainty product
$\Delta l \Delta p$ for various stationary barriers when the electrons
attempting to tunnel are initially in a momentum eigenstate. 

To summarize the procedure outlined in this section the steps in the
calculations will be the following. First we solve the
time-independent Schr\"odinger equation to find the electron wave
function in the presence of the barrier. From this we can calculate
the transmission and reflection coefficients and therefore $\Delta l$
with the use of Eq. (\ref{2.5}). Finally, we can find $\Delta p$ by using
the solution of Schr\"odinger equation and the potential $V_2(x)$ in
Eqs. (\ref{2.15})-(\ref{2.17}). 

\section{Uncertainty product for rectangular barriers}

In this section we use the formalism developed in Sec. II to calculate
the uncertainty product $\Delta l \Delta p$ for rectangular barriers,
both symmetric [see Fig. 2(a)] and asymmetric [see Fig. 2(b)] cases. 

In the symmetric barrier case the wave function which solves the
time-independent Schr\"odinger equation can be expresses as
\begin{eqnarray}
\psi_k(x)&=&\frac{1}{\sqrt{2\pi}} [e^{ikx}+r(k)e^{-ikx}], ~~~~~ x<a \nonumber \\
\psi_k(x)&=&\frac{1}{\sqrt{2\pi}} [C^+(k)e^{k_0x}+C^-(k)e^{-k_0x}], 
~~~~~ a<x<b \\
\psi_k(x)&=&\frac{1}{\sqrt{2\pi}} t(k)e^{ikx}, ~~~~~ x > b, \nonumber
\end{eqnarray}
where $\hbar k=\sqrt{2mE}$ and $\hbar k_0=\sqrt{2m(V_0-E)}$. The
normalization of the wave function used throughout this paper
corresponds to an incident flux $J_{in}=1/2\pi(\hbar k/m)$.

\begin{figure}[t]
\includegraphics[clip,width=0.20\columnwidth]{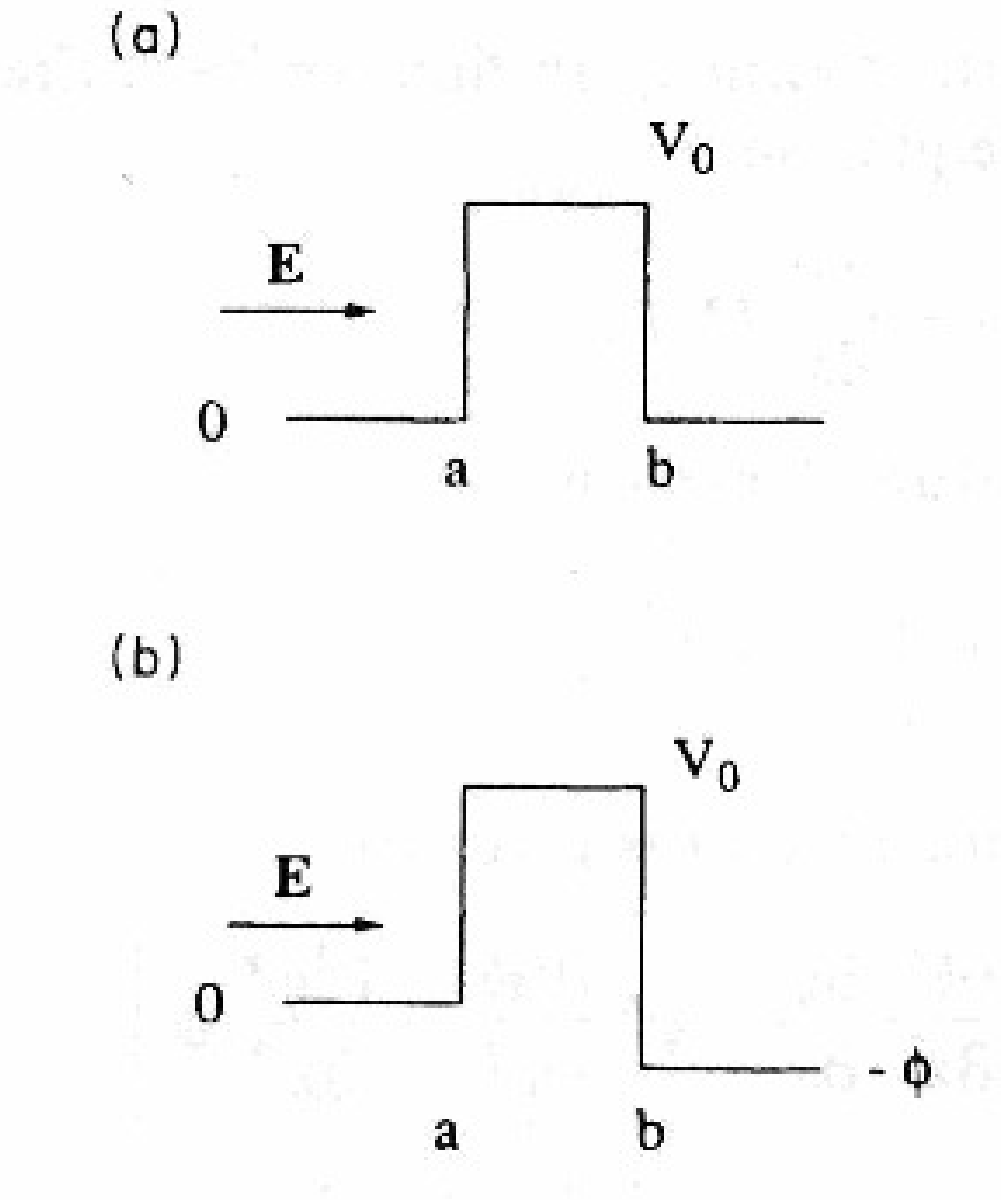}
\caption{Potential energy for a rectangular barrier; symmetric (a) and
  asymmetric (b).}
\label{fig2}
\end{figure}

By imposing the matching conditions for $\psi$ and the first
derivative of $\psi$ at $a$ and $b$ we obtain
\begin{equation}
t(k)=\frac{2 i k_0 k e^{-ik(b-a)}}{2 i k_0 k \cosh[k_0(b-a)]-(k_0^2-k^2)\sinh[k_0(b-a)]}.
\end{equation}
The uncertainty in the position $\Delta l$ can be calculated using
Eq. (\ref{2.5}): 
\begin{equation}
\Delta l= \frac{1}{\sqrt{N}} \frac{1}{T} \frac{k}{(k^2+k_0^2)\cosh(k_0l)},
\end{equation}
where $l=b-a$ and the transmission coefficient is
\begin{equation}
T=|t(k)|^2=\frac{1}{1+\frac{(k_0^2+k^2)\sinh^2[k_0(b-a)]}{4k_0^2 k^2}}.
\end{equation}
In order to calculate the momentum uncertainty we first define the
potential due to the test mass as
\begin{equation}
V_2(x)=V_0 \Theta(b-x).
\end{equation}
Using Eqs. (\ref{2.15}) and (\ref{2.16}) we then have
\begin{equation}
J_p^t=\frac{1}{2\pi} \frac{\hbar^2}{2m}(k^2-k_0^2)T,
\end{equation}
\begin{equation}
J_{p^2}^t=-\frac{1}{2\pi} \frac{\hbar^3}{m} k_0^2 k T.
\end{equation}
According to (\ref{2.17}), we get
\begin{equation}
(\Delta p)^2=N\frac{\hbar^2}{4 k^2} T [4 k^2 k_0^2+(k^2-k_0^2)^2T].
\end{equation}

On multiplying $\Delta l$ and $\Delta p$ we obtain $\Delta l \Delta
p=\hbar/2$, {\it i.e.}, the minimum uncertainty product for the test
mass. The uncertainty principle here may be regarded as arising from
the interaction with the tunneling electrons during the process of
measurement.

The same calculations can be repeated for an asymmetric rectangular
barrier as in Fig. 2(b), schematizing an unbiased barrier between two
materials having different work functions. The solution of the
Schr\"odinger equation is
\begin{eqnarray}
\psi_k(x)&=&\frac{1}{\sqrt{2\pi}} [e^{ikx}+r(k)e^{-ikx}], ~~~~ x<a \nonumber \\
\psi_k(x)&=&\frac{1}{\sqrt{2\pi}} [C^+(k)e^{k_0x}+C^-(k)e^{-k_0x}], 
~~~~ a \leq x<b \\
\psi_k(x)&=&\frac{1}{\sqrt{2\pi}} t(k)e^{i\bar{k}x}, ~~~~ x > b, \nonumber
\end{eqnarray} 
where $\hbar k=\sqrt{2mE}$, $\hbar k_0=\sqrt{2m(V_0-E)}$, and 
$\hbar \bar{k}=\sqrt{2m(E+\phi)}$. The transmission amplitude is found
to be
\begin{equation}
t(k)=\frac{2 i k_0 k e^{-i(\bar{k}b-ka)}}{i k_0 (k+\bar{k})
  \cosh[k_0(b-a)]+(k\bar{k}-k_0^2)\sinh[k_0(b-a)]}.
\end{equation}
and the transmission coefficient is now defined as
\begin{equation}
T=\frac{\bar{k}}{k}|t(k)|^2.
\end{equation}
To find the momentum uncertainty we can calculate the momentum flux
and the ``momentum-squared'' flux transmitted to the barrier via the
potential $V_2(x)=(V_0+\phi)\Theta(b-x)-\phi$ by applying (\ref{2.11})
and (\ref{2.14}):
\begin{equation}
J_p^t=\frac{1}{2\pi} \frac{\hbar^2}{2m}(\bar{k}^2-k_0^2)\frac{k}{\bar{k}}T,
\end{equation}
\begin{equation}
J_{p^2}^t=-\frac{1}{2\pi} \frac{\hbar^3}{m} k_0^2 k T.
\end{equation}
This allows us to evaluate $(\Delta p)^2$,
\begin{equation}
(\Delta p)^2=N\frac{\hbar^2}{4 \bar{k}^2} T [4 \bar{k}^2 k_0^2+(\bar{k}^2-k_0^2)^2T].
\end{equation}
It is straightforward to show numerically that in this case the
product of the uncertainties remains nearly the minimum allowed by
quantum mechanics, {\it i.e.}, $\hbar/2$, for any value of the
potential energy $\varphi$.

\section{Uncertainty product for a barrier with a constant electric field}
 
We now repeat the above procedure for a barrier in which a constant
electric field is present, such that the potential is expressed as
(see Fig. 3)

\begin{eqnarray}
V(x)&=& 0, ~~~~ x<a \nonumber \\
V(x)&=& , V_0-\phi \frac{x-a}{b-a}, ~~~~ a \leq x<b \\
V(x)&=& - \phi, ~~~~ x > b. \nonumber
\end{eqnarray}

\begin{figure}[t]
\includegraphics[clip,width=0.20\columnwidth]{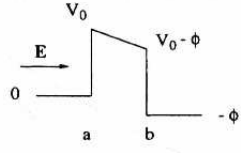}
\caption{Potential energy for a barrier with a constant electric field.}
\label{fig3}
\end{figure}

The solution of Schr\"odinger's time-independent equation is expressed
as 

\begin{eqnarray}
\psi_k(x)&=&\frac{1}{\sqrt{2\pi}} [e^{ikx}+r(k)e^{-ikx}], ~~~~ x<a \nonumber \\
\psi_k(x)&=&\frac{1}{\sqrt{2\pi}}
    [C^+(k)Ai[\alpha^{1/3}(\beta-x)]+C^-(k)Bi[\alpha^{1/3}(\beta-x)], ~~~~
      a \leq x<b \\
\psi_k(x)&=&\frac{1}{\sqrt{2\pi}} t(k)e^{i\bar{k}x}, ~~~~ x > b, \nonumber
\end{eqnarray} 
where $Ai(z)$ and $Bi(z)$ are the Airy functions with argument
$z=\alpha^{1/3}(\beta-x)$ and $\alpha=(2m/\hbar^2)\phi/(b-a)$, 
$\beta=a+[(V_0-E)/\phi](b-a)$. The quantities $k$ and $\bar{k}$ are
defined as in the previous case of the asymmetric rectangular barrier. 

After imposing the matching conditions we obtain

\begin{eqnarray}
r(k)&=&  -\frac{i}{2k} e^{ika}
 \{ik[C^+(k)Ai(\bar{a})+C^-(k)Bi(\bar{a})]+
\alpha^{1/3}[C^+(k)Ai'(\bar{a})+C^-(k) Bi'(\bar{a})]\}, \\
C^+(k)&=& \pi t(k) e^{\bar{k}b} [Bi'(\bar{b})+i \bar{k} \alpha^{-1/3}Ai(\bar{b})],\\
C^-(k)&=& -\pi t(k) e^{i\bar{k}b}[Ai'(\bar{b})+i\bar{k} \alpha^{-1/3} Ai(\bar{b})],
\end{eqnarray}
and
\begin{equation}
t(k)=-\frac{2ik}{\pi} e^{i(ka-\bar{k}b)} \alpha^{1/3}
\{[\alpha^{1/3}Ai'(\bar{a})-ik Ai(\bar{a})][\alpha^{1/3}Bi'(\bar{b})+i\bar{k} Bi(\bar{b})]
-[\alpha^{1/3}Bi'(\bar{a})-ik Bi(\bar{a})][\alpha^{1/3}Ai'(\bar{b})+i\bar{k} Ai(\bar{b})]\}^{-1},
\end{equation}
where
\begin{equation}
\bar{b}=\alpha^{1/3}(\beta-b)
\end{equation}
and 
\begin{equation}
\bar{a}=\alpha^{1/3}(\beta-a)
\end{equation}

\begin{figure}[t]
\includegraphics[clip,width=0.60\columnwidth]{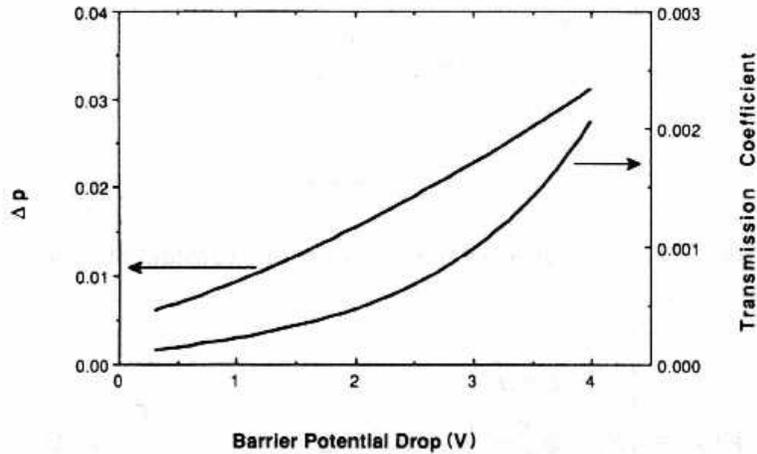}
\caption{Momentum uncertainty in units where $\hbar=1$ (left scale)
  and transmission coefficient (right scale) for the rectangular
  barrier {\it vs.} the applied voltage.}
\label{fig4}
\end{figure}
The prime denotes the derivative with respect to the argument of the
Airy function. The transmission and reflection coefficients are
defined as

\begin{equation}
T=\frac{\bar{k}}{k} |t(k)|^2, ~~~~ R=|r(k)|^2.
\end{equation}

We introduce, as in the previous considerations, a potential $V_2(x)$
defined as

\begin{equation}
V_2(x)=\left[V_0-\frac{\phi}{2}\right]\Theta(a-x)+\Theta(x-a)\Theta(b-x)
\left[V_0-\frac{\phi}{2}-\frac{\phi}{2} \frac{x-a}{b-a}\right]-
\phi \Theta(x-b).
\end{equation}
The momentum flux transmitted to the test mass can be calculated by

\begin{eqnarray}
J_p^t & = & J_p(b^+)+
\int_{a^+}^{b^-} \frac{\partial V_2}{\partial x}|\psi(x)|^2 dx +
\int_{b^-}^{b^+} \frac{\partial V_2}{\partial x}|\psi(x)|^2 dx =
\nonumber \\
& & J_p(b^+)+\frac{1}{2}[J_p(a^+)-J_p(b^-)]+[J_p(b^-)-J_p(B^+)]=\frac{1}{2}[J_p(a^+)+J_p(b^-)].
\end{eqnarray}
Analogously, for $J_{p^2}^t$ we obtain

\begin{equation}
J_{p^2}^t=\frac{1}{2}[J_{p^2}(a^+)+J_{p^2}(b^-)].
\end{equation}

Equations (\ref{2.15}) and (\ref{2.16}) can be used to express the fluxes inside
the barrier in terms of the fluxes calculated outside the barrier. The
detailed calculation of $\Delta p$ is shown in the Appendix, as well
as the explicit form of the derivative of the transmission coefficient
which allows us to obtain $\Delta l$.

In Fig. 4 the momentum uncertainty and the transmission coefficient
are shown as functions of the applied voltage. We observe that as the
potential drop across the barrier is increased the momentum
uncertainty increases. This can be understood in the following way:
when the electric field is increased the barrier is more transparent
to the electrons and can be effectively represented by a lower
rectangular barrier. Thus the electron current increases and $\Delta
l$ is reduced. The uncertainty relation therefore requires a larger
values of the momentum uncertainty. 

In Fig. 5 the uncertainty product versus the applied voltage is shown
for the case of $V_0=$5 eV and $E=$1 eV. The minimum uncertainty
product is obtained in the absence of an applied voltage which can
also be verified by examining the asymptotic behavior of the Airy
functions in the limit of $\phi \rightarrow 0$. The increase of the
uncertainty product $\Delta l \Delta p$ as the applied voltage is
raised is due, at least in part, to a correlation between $\Delta l$
and $\Delta p$ induced by the electric field in the gap. The mechanism
for the growth of the correlation is the following: an initial
momentum dispersion of the electrons will be transformed into a
spatial dispersion as the electrons traverse the barrier under the
influence of the electric field. The magnitude of the correlation has
been explicitly calculated in Ref. 8. A detailed discussion of the
correlation between the uncertainties in momentum and position may be
important for understanding techniques to surpass the standard quantum
limit by using time-dependent tunnel probe bias voltages. This is a
topic for further investigation.

\begin{figure}[t]
\includegraphics[clip,width=0.60\columnwidth]{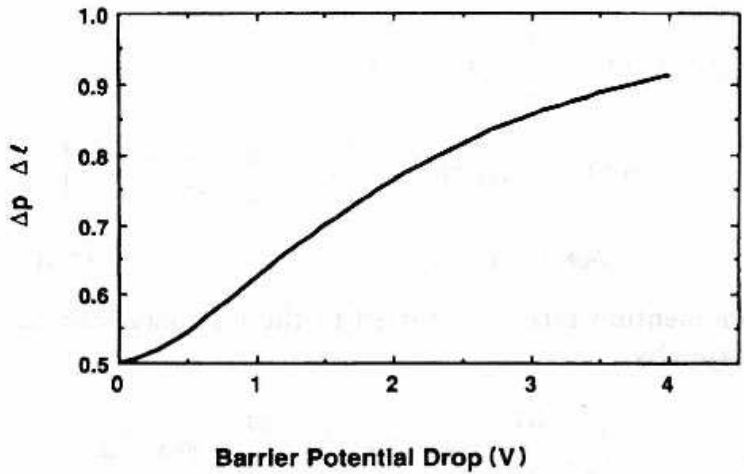}
\caption{Uncertainty product $\Delta l \Delta p$ in units $\hbar=1$
  for the test mass {\it vs.} the applied voltage in a rectangular
  barrier with a constant electric field. At a bias of 0 V, $\Delta l
  \Delta p=\hbar/2$, the value for a rectangular barrier.}
\label{fig5}
\end{figure}

\section{Practical configurations for reaching the quantum noise in
  tunneling transducers}

In this section we discuss some experimental aspects of quantum
measurements with a tunneling transducer. To compare the quantum noise
with the classical sources of noise we have to introduce a ``quantum''
force noise spectral density which allows us to use the usual
techniques of stochastic processes \cite{PRB9}. We do not claim to
rigorously define such an effective noise spectral density, although
it may be possible to define such a tool in the framework of Nelson's
stochastic mechanics \cite{PRB10}. Following Ref. 8 we write the
spectral density of the force fluctuations in terms of the variance of
the momentum current per unit bandwidth
\begin{equation}
S_{f_Q}=2 \frac{(\Delta p)^2}{\tau},
\label{5.1}
\end{equation}
where the effective force spectral density for the quantum noise has
been expressed in terms of the uncertainty in the momentum deposited
in the test mass by the tunneling current $I_0$ in a time interval
$\tau$. In practical cases the bias voltage applied to the tunnel
junction will be much less than the work function of the tip material so
the rectangular barrier is a close approximation to reality. In this
case the spectral density of the ``quantum'' force noise is
\begin{equation}
S_{f_Q}=\frac{I_0}{e} \hbar^2 k^2 \frac{1}{2}
\left[\left(1+\left(\frac{k_0}{k}\right)^2\right)^2- 
      \left(1-\left(\frac{k_0}{k}\right)^2\right)^2(1-T) \right]. 
\end{equation}
The biggest practical obstacle to observing the quantum effects we
discuss in this paper is the thermal noise which is manifested as the
Brownian motion of the test mass. One can describe the Brownian motion
of the test mass by including a Langevin force having a single-sided
spectral density
\begin{equation}
S_{f_L}=4m (2 \pi f_0) \frac{k_B \theta}{Q}.
\label{5.3}
\end{equation} 
We assumed in Eq. (\ref{5.3}) that the test mass is a mechanical resonator,
at a temperature $\theta$, having mass $m$, frequency $f_0$, and
quality factor $Q$ such that the decay time of the free oscillation is
$Q/\pi f_0$. To be able to observe the influence of the tunneling
transducer on the test mass the Langevin force must be smaller than
the force fluctuations from the tunneling transducer, {\it i.e.}, 
$S_{f_Q} > S_{f_L}$. This can be expressed in the following practical
form:

\begin{equation}
\frac{10^{-6} {\mathrm A }}{I_0} \frac{m}{10^{-10} {\mathrm kg}}\frac{\theta}{10 {\mathrm mK}} 
\frac{f_0}{10^5 {\mathrm Hz}} \frac{10^7}{Q} < 1.
\label{5.4}
\end{equation}

\begin{figure}[b]
\includegraphics[clip,width=0.80\columnwidth]{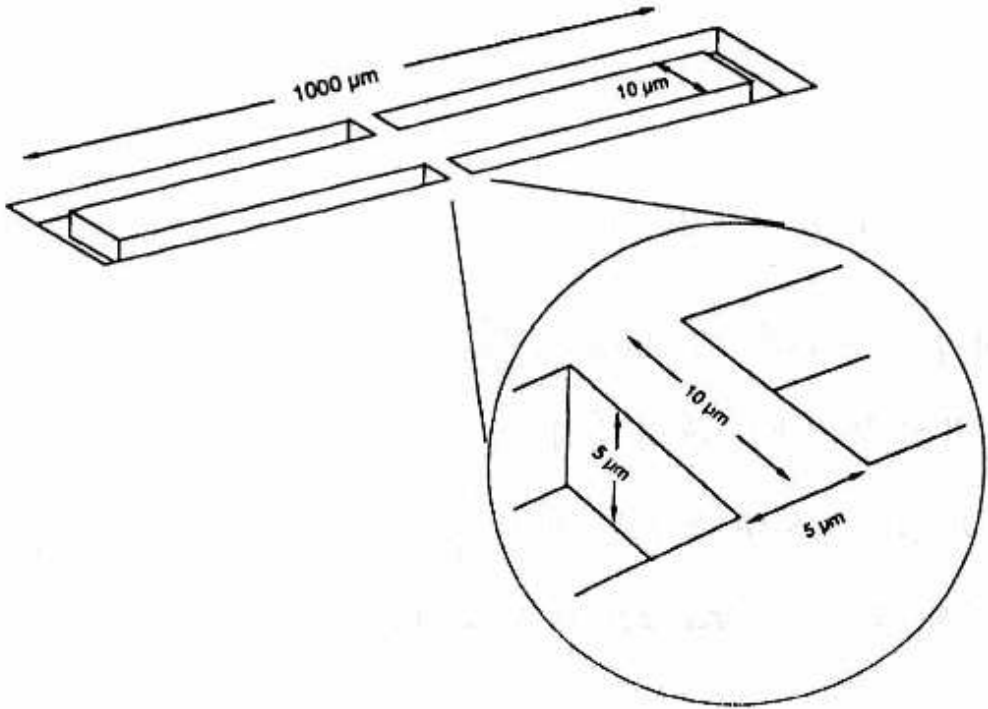}
\caption{Design for a silicon micromachined torsional resonator with a
  resonant frequency of 60 KHz and an effective mass of $10^{-10}$ kg.}
\label{fig6}
\end{figure}

We have assumed that $k_0=10^{10} {\mathrm m}^{-1}$. The mass,
frequency, and $Q$ used in (\ref{5.4}) are appropriate to micromachined 
silicon resonators at low temperatures. The mass of $10^{-10}$ kg
which is assumed above corresponds to a silicon structure like the one
shown in Fig. 6. A mechanical quality factor of $6 \times 10^5$ was
obtained at room temperature in a micromachined silicon torsional
resonator of mass $7 \times 10^{-6}$ kg \cite{PRB11}, and a more
massive resonator, $m \simeq 1$ g, which had a similar $Q$ at room
temperature achieved a $Q$ approaching $10^8$ at 10 mK \cite{PRB12}. 
A systematic study of acoustic losses of silicon resonators at
cryogenic temperatures indicates that the intrinsic $Q's$ of silicon 
are over one billion \cite{PRB13}. Therefore we assume that a $Q$ of
$10^7$ may be achievable with a $10^{-10}$ kg mechanical resonator at
10 mK. Atomic force-sensing microcantilevers in this mass range have
been fabricated and used at room temperature \cite{PRB14}.

There are two practical problems which are somewhat eased by working
with a mechanical resonator at a frequency of 100 kHz. The first is
the $1/f$ noise in the tunneling current. At a frequency of 100 kHz it
is likely that the $1/f$ noise component in the tunneling current
should be below the level of the shot noise. Furthermore there should
be no problems with seismic vibrations and vibration isolation at the
frequency of 100 kHz. Two other requirements to reach the quantum
noise limit with the tunneling transducer are that the noise of the
preamplifier used to sense the tunnel current be small in comparison
to the shot noise and that the dynamic capacitance of the tunnel probe
also be small. The first requirement, that the amplifier noise be
insignificant compared to the tunneling current shot noise, is fairly
easy to meet. For a tunneling current of $10^{-6}$ A the shot noise
spectral density is $5.7 \times 10^{-13}$ A/$\sqrt{\mathrm Hz}$. This
is a fairly high noise level compared to the noise of commonly
available transistors and operational amplifiers \cite{PRB15}. The
other requirement is that the dynamical capacitance, {\it i.e.}, the
probe capacitance which changes as the inverse of the tip to test mass
gap, be ``small''. Small in this context means low enough to ensure
that the back-action force associated with fluctuations in the energy
stored in the capacitor is less than the quantum force fluctuations of
the tunneling current. The specific requirement on the dynamic
capacitance have been calculated and values of $10^{-17}$ F or smaller
are needed to reach the quantum limit \cite{PRB7}. There is also some
experimental evidence that the dynamic capacitance of tunnel probes
can be in this range \cite{PRB16}. Note that the stray probe
capacitance, which may be orders of magnitude larger, is not important
in this respect because it is only weakly gap dependent.

In the realm of conventional, nontunneling transducers, to overcome
the two problems just discussed the so-called back-action evasion
(BAE) techniques have been developed \cite{PRB2}. One example of a BAE
strategy which has been used on a capacitive transducer coupled to a
mechanical harmonic oscillator is to perform phase-sensitive
detection. The coupling, {\it i.e.}, the electric field in the
capacitor formed between the transducer and the test mass, is
modulated, which has the consequence that the back-action force acts
on one of the phases of the mechanical oscillator while the
information which is extracted from the mechanical oscillator reflects
the state of the orthogonal phase. It is not clear if one can directly
apply the phase-sensitive continuous-monitoring BAE techniques to the
tunneling transducer and circumvent the quantum limit, however, one
may be able to use a so-called quantum nondemolition stroboscopic
measurement \cite{PRB2}. In this way, provided that an initial
high-sensitivity measurement of the position has been made, repeated
measurements made at time intervals of half the period of the
mechanical oscillator motion can be performed with the same accuracy
as the initial precise measurement. A stroboscopic measurement could
be realized by sending in short pulses of tunneling current at time
intervals equal to one-half of the period of the mechanical
oscillator. One needs very short pulses to make an accurate
stroboscopic measurement \cite{PRB17}. The duration of each pulse, and
therefore the accuracy of the stroboscopic measurement, will be
limited by the RC time constant of the tunneling probe so one will
have to avoid large stray capacitance of the tunneling probe.

\section{Conclusions}

Quantum-mechanical uncertainties for measurements made on a test mass
by a tunneling transducer have been calculated using a
first-quantization approach. 

The possibility of reaching the quantum limit in practical
electromechanical devices incorporating a tunneling probe has been
discussed, as well as a possible way to surpass the standard quantum
limit by means of quantum nondemolition stroboscopic techniques.

in closing the ability to probe a single macroscopic object in the
quantum domain opens a fundamental question concerning the validity of
the quantum ergodicity assumption. This assumption is that ensemble
averaged quantities are equivalent to time averages of the same
quantity in a single quantum system. The point in our analysis where
the quantum ergodicity assumption enters is Eq. (\ref{5.1}), in which we
assert the equivalence of the quantum uncertainties we calculated and
the noise spectral densities of the corresponding quantities. All the
experiments which probe microscopic quantum phenomena in which a
measurement is made on each member of an ensemble of identically
prepared systems have outcomes, without exception, which agree with
the predictions of quantum mechanics. The sort of experiments which
should be possible with the tunneling transducer are qualitatively
different. One will be able to make repeated measurements on a single
quantum system which is weakly coupled to its environment. In this
case the outcome of the later measurements will depend upon the
interaction of the measuring apparatus with the system during earlier
measurements. If the ergodic assumption is true, then the outcome of
the experiments discussed in this paper should coincide with the
quantum predictions obtained by the usual ensemble-averaging technique.

\acknowledgments
We thank G. Jona-Lasinio, F. Marchesoni, F. Sacchetti, and R. Koch for
their helpful comments. R.O. would like to thank INFN-Sezione di Roma
for financial support. M.B. would like to acknowledge support under a
grant from the Office of Naval Research.

\vspace{1.0cm}
\noindent
\centerline{\bf APPENDIX}

\vspace{0.8cm}
We use (\ref{2.15}) and (\ref{2.16}) for expressing the momentum density flux and
the square momentum density flux inside the barrier in terms of the
analogous quantities outside the barrier, 

\begin{eqnarray}
  J_p(b^-) & = & J_p(b^+)- \frac{|t(k)|^2}{2\pi}V_0, \\
  J_{p^2}(b^-) & = & J_{p^2}(b^+)-\frac{|t(k)|^2}{2\pi}2 V_0 \hbar \bar{k},\\
  J_p(a^+) & = & J_p(a^-)- \frac{V_0}{2\pi}[1+R+2 Re(r e^{-2ika})], \\
  J_{p^2}(a^+) & = & J_{p^2}(a^-)-\frac{V_0}{2\pi} 2 \hbar k(1-R),
\end{eqnarray}
where the currents outside the barrier are
\begin{eqnarray}
  J_p(b^+) & = & \frac{|t(k)|^2}{2\pi} \frac{\hbar^2 \bar{k}^2}{m}, \\
  J_{p^2}(b^-) & = & \frac{|t(k)|^2}{2\pi} \frac{\hbar^3 \bar{k}^3}{m},\\
  J_p(a^+) & = & \frac{1+R}{2\pi}\frac{\hbar^2 k^2}{m}, \\
  J_{p^2}(a^+) & = & \frac{1-R}{2\pi} \frac{\hbar^3 k^3}{m},
\end{eqnarray}
which, using (\ref{2.17}), allows us to find $(\Delta p)^2$.

Finally the first derivative of the transmission coefficient with
respect to the gap of the tunneling probe can be calculated and we
find $\Delta l$ using (\ref{2.5}):

\begin{equation}
  \frac{dT}{dl}=\frac{\bar{k}}{k}\frac{d}{dl}|t(k)|^2=\frac{\bar{k}}{k} \left[\frac{dt(k)}{dl} t^*(k)+t(k)\frac{dt(k)^*}{dl} \right],
\end{equation}
where the derivative of $t(k)$ is evaluated as

\begin{align}
  \frac{dt(k)}{dl} =&
  t(k)\left[\alpha^{-1/3}\frac{d\alpha^{1/3}}{dl}-i(k+\bar{k})\right]
  +t(k)^2 \frac{\pi}{2ik}\alpha^{-1/3}e^{i(\bar{k}b-ka)} 
  \nonumber \\ 
  &\times  \biggl[ \left\{
    [\alpha^{1/3} \bar{a} Ai(\bar{a})-ikAi'(\bar{a})]
    [\alpha^{1/3}Bi'(\bar{b})+i\bar{k}Bi(\bar{b})] \right.
  \nonumber \\ &\qquad 
  \left. -[\alpha^{1/3} \bar{a} Bi(\bar{a})-ikBi'(\bar{a})]
    [\alpha^{1/3}Ai'(\bar{b})+i\bar{k}Ai(\bar{b})] \right\} 
  \frac{d\bar{a}}{dl} 
  \nonumber \\ &\quad
  +\left\{ 
    [\alpha^{1/3}Ai'(\bar{a})-ikAi(\bar{a})]
    [\alpha^{1/3}\bar{b}Bi(\bar{b})+i\bar{k}Bi'(\bar{b})] \right. 
  \nonumber \\  &\qquad
  \left. -[\alpha^{1/3} Bi'(\bar{a})-ikBi(\bar{a})]
    [\alpha^{1/3}\bar{b}Ai(\bar{b})+i\bar{k}Ai'(\bar{b})] \right\}
  \frac{d\bar{b}}{dl} 
  \nonumber \\ &\quad
  +\left\{ [\alpha^{1/3} Bi'(\bar{b})+i\bar{k}Bi(\bar{b})]Ai'(\bar{a})
    +[\alpha^{1/3}Ai'(\bar{a})-ik Ai(\bar{a})] Bi'(\bar{b}) \right. 
  \nonumber \\ &\qquad 
  \left. -[\alpha^{1/3} Ai'(\bar{b})+i\bar{k}Ai(\bar{b})]Bi'(\bar{a})
    -[\alpha^{1/3}Bi'(\bar{a})-ik Bi(\bar{a})] Ai'(\bar{b})\right\}
  \frac{d\alpha^{1/3}}{dl} \biggr],
  \end{align}
  taking into account the following relationships:
\begin{eqnarray}
  \frac{dt^*(k)}{dl}  & = & \left[\frac{dt(k)}{dl}\right]^* ,\\
  \frac{d\bar{a}}{dl} & = & \frac{2}{3} \left[\frac{2m\phi}{\hbar^2}\right]^{1/3} \frac{V_0-E}{\phi} l^{-1/3},  \\
  \frac{d\bar{b}}{dl} & = & \frac{2}{3} \left[\frac{2m\phi}{\hbar^2}\right]^{1/3} \left[\frac{V_0-E}{\phi}-1\right]{\varphi} l^{-1/3},  \\
  \frac{d\alpha^{1/3}}{dl} & = & -\frac{1}{3} \frac{\alpha^{1/3}}{l}.
\end{eqnarray}

\end{document}